\documentclass[10pt,letterpaper,twocolumn]{article} 

\usepackage{ol2}
\usepackage{amsmath}
\providecommand{\abs}[1]{\left\lvert#1\right\rvert}

\begin{document}

\twocolumn[ 

\title{Influence of atmospheric turbulence on states of light carrying orbital angular momentum}

\author{Brandon Rodenburg,$^{1}$$^*$ Martin P. J. Lavery,$^2$$^{\dagger}$ Mehul Malik,$^{1}$ Malcolm N. O'Sullivan,$^{1}$ Mohammad Mirhosseini,$^{1}$ David~J.~Robertson$^{3}$, Miles Padgett,$^2$ and Robert W. Boyd$^{1,4}$}

\address{
$^1$The Institute of Optics, University of Rochester. 320 Wilmot BLDG, 275 Hutchison Rd, Rochester NY 14627, USA
\\
$^2$School of Physics and Astronomy, University of Glasgow. SUPA, Kelvin Building, Glasgow G12 8QQ, Scotland, UK
\\
$^3$Centre for Advanced Instrumentation, Department of Physics, Durham University, Durham, DH1 3LE, UK
\\
$^4$Department of Physics, University of Ottawa, Ottawa, ON K1N 6N5, Canada
\\
$^*$Brandon.Rodenburg@gmail.com 

$^{\dagger}$m.lavery@physics.gla.ac.uk
}

\begin{abstract}
We have experimentally studied the degradation of mode purity for light beams carrying orbital angular momentum (OAM) propagating through simulated atmospheric turbulence. The turbulence is modeled as a randomly varying phase aberration, which obeys statistics postulated by Kolmogorov turbulence theory. We introduce this simulated turbulence through the use of a phase-only spatial light modulator. Once the turbulence is introduced, the degradation in mode quality results in cross-talk between OAM modes. We study this cross-talk in OAM for eleven modes, showing that turbulence uniformly degrades the purity of all the modes within this range, irrespective of mode number.
\end{abstract}

\ocis{010.1330, 270.5585, 270.5565}

 ] 
\noindent
A fundamental concern for any free-space communications channel is the effect that atmospheric turbulence has on the cross-talk between channels. Atmospheric turbulence has been studied at great length by the astronomy community in relation to aberrations in an image \cite{Beckers1993}. The natural randomly time dependent variations in temperature and pressure of the atmosphere resulting in a change in density of the atmosphere result in a spatial dependent change of the refractive index leading to a phase distortion across a transmitted beam \cite{tatarski1961}. A phase distortion of this type can be considered as a phase screen, and is commonly referred to as thin phase turbulence \cite{Fried1965}.

There has been recent interest in the use of spatial modes as an additional degree of freedom to increase the available information bandwidth for free-space communication. One example of these modes are beams carrying orbital angular momentum (OAM). Allen \emph{et~al.} showed that beams with a transverse amplitude profile of $\psi_\ell=A(r) \exp(i \ell \theta)$ carry an orbital angular momentum of $\ell \hbar$ per photon \cite{Allen1992}. An example of such beams are Laguerre-Gaussian (LG) modes which have a helical phase structure, with $r$ and $\theta$ as the radial and angular coordinates respectively. The variable $\ell$ is an unbounded integer, and as such suggests the use of OAM as a variable in free-space optical communication links \cite{Gibson2004,Tamburini2012}. In addition to the advantages of a large alphabet, the security of cryptographic keys transmitted with a quantum key distribution system have been shown to be improved with the use of a large Hilbert space \cite{Bourennane2001}.

\begin{figure}
    \centerline{\includegraphics[width=6.6cm]{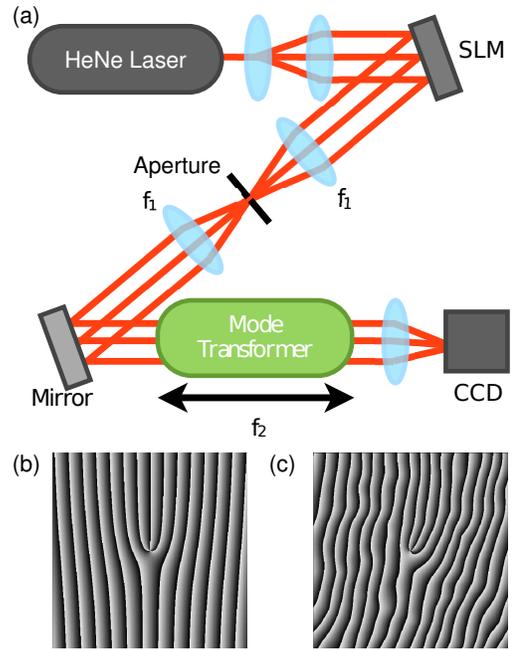}}
    \caption{(a) A beam carrying OAM is prepared by the use of a $\ell$-forked hologram, seen in (b). This is realized on a spatial light modulator (SLM) illuminated by an expanded HeNe laser. The first order beam is imaged onto the front aperture of a OAM mode sorter (MS) which converts OAM states into transverse momentum states with the use of two refractive optical elements. These transverse momentum states are then focused to specific spatial locations on a CCD. The power measured in each of these locations gives a measure of the OAM superposition incident on the mode sorter. (c) Thin phase turbulence is added to the $\ell$-forked hologram changing the OAM superposition measured by the system.}
    \label{ExpSetup}
\end{figure}

Recently, there have been several studies on how atmospheric turbulence affects such OAM based communication \cite{Paterson2005,Tyler2009,Smith2006,Gbur2008,Roux2011}. In this letter, we experimentally study the effects of atmospheric turbulence on a communication system utilizing OAM modes as the information carrier. We generate a single OAM mode using a spatial light modulator (SLM). Atmospheric turbulence is then simulated by the addition of a turbulent phase screen to the phase hologram displayed on the SLM shown in Fig.~\ref{ExpSetup}. Once the turbulence is applied, the phase aberrations result in a spread of the input mode power over neighboring OAM modes, resulting in cross-talk between the channels. This spread in power is then measured for different turbulence strengths. 

We generate turbulence phase screens according to Kolmogorov turbulence theory \cite{Fried1965}. The aberrations introduced by atmospheric turbulence can be considered as normal random variables, where the ensemble average can be written as $ \left<\left[\phi(\mathbf{r_1})-\phi(\mathbf{r_2})\right]^2\right>$, which is known as the phase structure function \cite{Paterson2005,Tyler2009}. Here, $\phi(\mathbf{r_1})$ and $\phi(\mathbf{r_2})$ are two randomly generated phase fluctuations. From Kolmogorov statistics it can be shown that this ensemble average must meet the requirement that
\begin{equation}
    \left<\left[\phi(\mathbf{r_1})-\phi(\mathbf{r_2})\right]^2\right> = 6.88\abs{\frac{\mathbf{r}_1-\mathbf{r}_2}{r_0}}^{5/3}.
\end{equation}
The value $r_0$ is the Fried parameter, and is a measure of the traverse distance scale over which the refractive index is correlated \cite{Fried1965}. To characterize the effect of turbulence on the optical system, the ratio $D/r_0$ is considered, where $D$ is the aperture of the system. There are two limiting cases for this ratio: when $D/r_0 < 1$, the resolution of the system is limited by its aperture, and when $D/r_0 > 1$, the atmosphere limits the system's ability to resolve an object \cite{Fried1965}.

Our theoretical analysis closely follows that of reference \cite{Tyler2009}. Consider a single OAM mode,  $\psi_{\ell}$, transmitted through an ensemble average of many turbulent phase screens. The average detected power, $s_\Delta$, in the mode, $\psi_{\ell+\Delta}$, is given by
\begin{equation}
    s_\Delta =
    \frac{1}{\pi}\int_0^1\! \rho \,\mathrm{d}\rho \!
    \int_0^{2\pi}\! \mathrm{d}\theta \, e^{-3.44\left[\left(\frac{D}{r_0}\right)\left(\rho\sin{\frac{\theta}{2}}\right)\right]^{5/3}}\cos{\Delta\theta},
    \label{Prediction}
\end{equation}
where $\Delta$ is an integer step in the mode index of $\ell$, and $\rho=2r/D$ \cite{Tyler2009}.

As shown in Fig. 1, we generate OAM modes by use of a simple forked diffraction grating created using an SLM that is illuminated by an expanded gaussian beam produced by a HeNe laser.  Rather than producing a pure Laguerre-Gaussian mode, this results in a helically phased beam, which has a near-Gaussian intensity distribution in the image plane of the SLM.  This approach maintains the ratio $D/r_0$ independent of the mode index. A particular turbulent phase screen can then be added to this hologram to simulate the presence of atmospheric turbulence. The SLM is then imaged to the $8$ mm diameter input pupil of the OAM mode sorter (MS) to decompose the resulting beam into its constituent OAM modes.

The mode sorter uses two refractive optical elements which transform OAM states into transverse momentum states \cite{Berkhout2010,Lavery2012}. These elements transform a beam of the form $\exp(i\ell \theta)$, to $\exp(i \ell x/a)$ at the output, where $a$ is a scaling parameter. A lens is used to focus these transverse momentum modes to discrete spots at a CCD placed in its focal plane. Adjacent, equally sized regions are selected on the CCD image, with each region corresponding to a specific $\ell$-value. The total counts over all the pixels in each region is summed to give the relative power in each OAM mode. For each input $\ell$ mode, the power is measured across all 11 regions, and normalized with respect to the power measured for $\ell=0$ with no turbulence applied. 

\begin{figure}
    \def\svgwidth{8.5cm}
    \begingroup%
      \makeatletter%
      \providecommand\color[2][]{%
        \errmessage{(Inkscape) Color is used for the text in Inkscape, but the package 'color.sty' is not loaded}%
        \renewcommand\color[2][]{}%
      }%
      \providecommand\transparent[1]{%
        \errmessage{(Inkscape) Transparency is used (non-zero) for the text in Inkscape, but the package 'transparent.sty' is not loaded}%
        \renewcommand\transparent[1]{}%
      }%
      \providecommand\rotatebox[2]{#2}%
      \ifx\svgwidth\undefined%
        \setlength{\unitlength}{273.75707296bp}%
        \ifx\svgscale\undefined%
          \relax%
        \else%
          \setlength{\unitlength}{\unitlength * \real{\svgscale}}%
        \fi%
      \else%
        \setlength{\unitlength}{\svgwidth}%
      \fi%
      \global\let\svgwidth\undefined%
      \global\let\svgscale\undefined%
      \makeatother%
      \begin{picture}(1,0.71916627)%
        \put(0,0){\includegraphics[width=\unitlength]{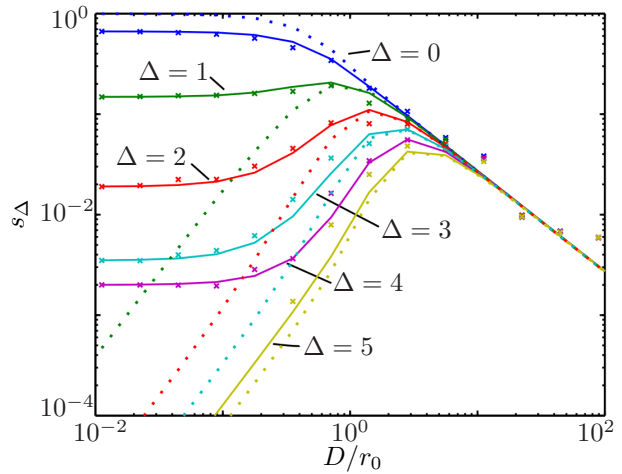}}%
        \put(0.06461806,0.69216423){\makebox(0,0)[lb]{\smash{$10^0$}}}%
        \put(0.02220263,0.37688109){\color[rgb]{0,0,0}\rotatebox{90}{\makebox(0,0)[lb]{\smash{$s_\Delta$}}}}%
        \put(0.04705002,0.39006482){\makebox(0,0)[lb]{\smash{$10^{-2}$}}}%
        \put(0.04705002,0.08427845){\makebox(0,0)[lb]{\smash{$10^{-4}$}}}%
        \put(0.50242364,0.04507091){\makebox(0,0)[lb]{\smash{$10^0$}}}%
        \put(0.49045002,0.00827033){\color[rgb]{0,0,0}\makebox(0,0)[lb]{\smash{$D/r_0$}}}%
        \put(0.09928282,0.04507091){\makebox(0,0)[lb]{\smash{$10^{-2}$}}}%
        \put(0.88731688,0.04507091){\makebox(0,0)[lb]{\smash{$10^2$}}}%
        \put(0.16581343,0.47782233){\makebox(0,0)[lb]{\smash{$\Delta=2$}}}%
        \put(0.20131747,0.61114396){\makebox(0,0)[lb]{\smash{$\Delta=1$}}}%
        \put(0.56396187,0.64092802){\makebox(0,0)[lb]{\smash{$\Delta=0$}}}%
        \put(0.58067677,0.36380159){\makebox(0,0)[lb]{\smash{$\Delta=3$}}}%
        \put(0.4586642,0.1803873){\makebox(0,0)[lb]{\smash{$\Delta=5$}}}%
        \put(0.50516273,0.28062846){\makebox(0,0)[lb]{\smash{$\Delta=4$}}}%
      \end{picture}%
    \endgroup%
    \caption{The average power ($s_\Delta$) in detected mode $\psi_{\Delta}$ is plotted as a function of turbulence strength ($D/r_0$) for an input mode with $\ell=0$ (see Eqn. \ref{Prediction}). Experimental data (crosses) is co-plotted with the theoretical prediction given by Eqn. \ref{Prediction} taking into account the inherent cross-talk of the sorter (solid lines). The original theory from Ref. \cite{Tyler2009} is also plotted for comparison (dotted lines).}
    \label{Results}
\end{figure}

A mode range of $\ell = -5$ to $\ell= +5$ was investigated, and for $100$ randomly generated phase screens the average power in each OAM mode was measured (Fig.~\ref{Results}). A range of turbulence levels characterised by $D/r_0$ were tested.  As predicted by Eqn. \ref{Prediction}, the cross-talk between OAM modes increases with turbulence. In the mid/high turbulence regime we see good agreement between our measurements and the theory proposed in \cite{Tyler2009}. In the low turbulence regime, the cross-talk between modes arises from residual cross-talk in our mode sorter, which can be attributed to the diffraction limit \cite{Berkhout2010,Lavery2012}. The weightings of the known input states described by an $N=11$ element column vector $[I]$ are mapped by an $N \times N$ cross-talk matrix onto the measured $N$ element output vector $[O]$ (Eqn. \ref{crosstalk}). For the case of zero residual cross-talk, this matrix would correspond to an identity matrix. For finite cross-talk, the coefficients $a-j$ etc. are measured at zero turbulence. Consequently, this matrix is used to predict the measured OAM output spectrum for an input OAM state subject to the atmospheric cross-talk from our theoretical model (Eqn. \ref{Prediction}).

\begin{equation}
    \begin{bmatrix} O_0 \\ O_1 \\ \vdots\\O_N \end{bmatrix} =
    \begin{bmatrix}1-g & a & \hdots & b \\c & 1-h & \hdots  & d \\ \hdots  & \hdots  & \hdots  & \hdots  \\e & f & \hdots  & 1-f\end{bmatrix}
    \begin{bmatrix} I_0 \\ I_1 \\ \vdots\\I_N \end{bmatrix}
    \label{crosstalk}
\end{equation}

It is seen in Fig.~\ref{Results}, that at high turbulence values ($D/r_0 \gg 1$) the average power is equally spread between all detected modes. It should be noted that we are only considering the proportion of the power detected within the detector regions and not considering the power incident outside these regions.
\begin{figure}
    \def\svgwidth{8.5cm}
    \begingroup%
      \makeatletter%
      \providecommand\color[2][]{%
        \errmessage{(Inkscape) Color is used for the text in Inkscape, but the package 'color.sty' is not loaded}%
        \renewcommand\color[2][]{}%
      }%
      \providecommand\transparent[1]{%
        \errmessage{(Inkscape) Transparency is used (non-zero) for the text in Inkscape, but the package 'transparent.sty' is not loaded}%
        \renewcommand\transparent[1]{}%
      }%
      \providecommand\rotatebox[2]{#2}%
      \ifx\svgwidth\undefined%
        \setlength{\unitlength}{279.03704718bp}%
        \ifx\svgscale\undefined%
          \relax%
        \else%
          \setlength{\unitlength}{\unitlength * \real{\svgscale}}%
        \fi%
      \else%
        \setlength{\unitlength}{\svgwidth}%
      \fi%
      \global\let\svgwidth\undefined%
      \global\let\svgscale\undefined%
      \makeatother%
      \begin{picture}(1,1.20974265)%
        \put(0,0){\includegraphics[width=\unitlength]{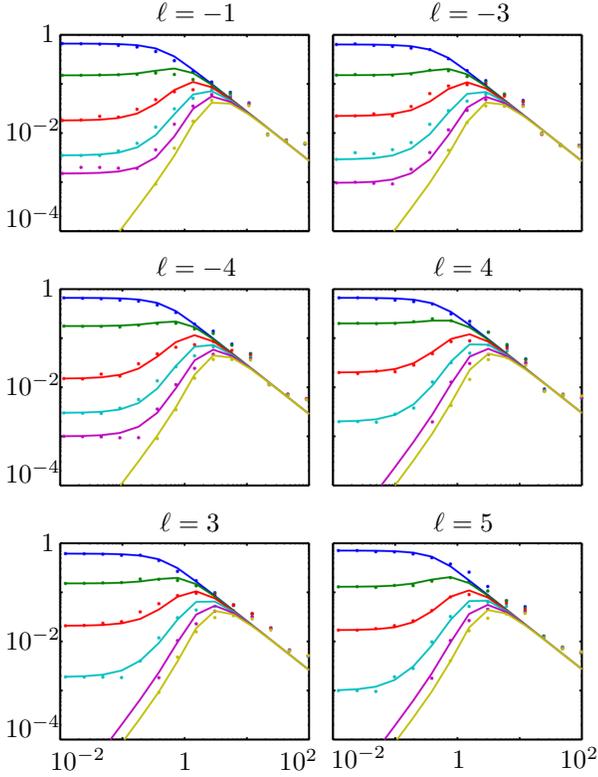}}%
        \put(0.23309518,1.17723765){\color[rgb]{0,0,0}\makebox(0,0)[lb]{\smash{$\ell =-1$}}}%
        \put(0.23309518,0.7785898){\color[rgb]{0,0,0}\makebox(0,0)[lb]{\smash{$\ell =-4$}}}%
        \put(0.23931076,0.3799416){\color[rgb]{0,0,0}\makebox(0,0)[lb]{\smash{$\ell =3$}}}%
        \put(0.06717541,0.00561081){\color[rgb]{0,0,0}\makebox(0,0)[lb]{\smash{$10^{-2}$}}}%
        \put(0.26793601,0.00533368){\color[rgb]{0,0,0}\makebox(0,0)[lb]{\smash{$1$}}}%
        \put(0.42527721,0.00533368){\color[rgb]{0,0,0}\makebox(0,0)[lb]{\smash{$10^2$}}}%
        \put(0.0535083,1.14518816){\color[rgb]{0,0,0}\makebox(0,0)[lb]{\smash{$1$}}}%
        \put(-0.0028558,0.99151056){\color[rgb]{0,0,0}\makebox(0,0)[lb]{\smash{$10^{-2}$}}}%
        \put(-0.0028558,0.85475785){\color[rgb]{0,0,0}\makebox(0,0)[lb]{\smash{$10^{-4}$}}}%
        \put(0.0535083,0.74654021){\color[rgb]{0,0,0}\makebox(0,0)[lb]{\smash{$1$}}}%
        \put(-0.0028558,0.59286261){\color[rgb]{0,0,0}\makebox(0,0)[lb]{\smash{$10^{-2}$}}}%
        \put(-0.0028558,0.45610978){\color[rgb]{0,0,0}\makebox(0,0)[lb]{\smash{$10^{-4}$}}}%
        \put(0.0535083,0.3478922){\color[rgb]{0,0,0}\makebox(0,0)[lb]{\smash{$1$}}}%
        \put(-0.0028558,0.19421458){\color[rgb]{0,0,0}\makebox(0,0)[lb]{\smash{$10^{-2}$}}}%
        \put(-0.0028558,0.05746181){\color[rgb]{0,0,0}\makebox(0,0)[lb]{\smash{$10^{-4}$}}}%
        \put(0.66086736,1.17723765){\color[rgb]{0,0,0}\makebox(0,0)[lb]{\smash{$\ell =-3$}}}%
        \put(0.66708293,0.7785898){\color[rgb]{0,0,0}\makebox(0,0)[lb]{\smash{$\ell =4$}}}%
        \put(0.66708293,0.3799416){\color[rgb]{0,0,0}\makebox(0,0)[lb]{\smash{$\ell =5$}}}%
        \put(0.50641554,0.00847782){\color[rgb]{0,0,0}\makebox(0,0)[lb]{\smash{$10^{-2}$}}}%
        \put(0.69570812,0.00820068){\color[rgb]{0,0,0}\makebox(0,0)[lb]{\smash{$1$}}}%
        \put(0.86451731,0.00820068){\color[rgb]{0,0,0}\makebox(0,0)[lb]{\smash{$10^2$}}}%
      \end{picture}%
    \endgroup
    \caption{The spread in power resulting from atmospheric turbulence was measured for a range of different propagating OAM modes $\psi_{\ell}$. }
    \label{results2}
\end{figure}

The theory presented in Ref. \cite{Tyler2009} indicates that the probability of modal cross-talk resulting from atmospheric turbulence is independent of the input mode number. To examine this theory, we studied the effects of turbulence on different OAM modes ranging from $\ell = -5$ to $\ell= +5$. For each of these modes, the same set of turbulent phase screens was applied. The measured cross-talk is shown in Fig.~\ref{results2}.  We note that the observed cross-talk is indeed very similar for the entire range of OAM modes that we examined.  

In this work we have studied the case where turbulence can be considered as a thin phase screen. Such an approach is widely used in astronomy, as when one considers the distance to an astronomical light source, the largest proportion of the turbulence is experienced relatively close to the observer. However, in the case of long distance point-to-point communications on earth, turbulence is characterized more accurately by multi-plane turbulence. In such cases one can expect intensity fluctuations and scintillation effects, and the thin phase model is insufficient.  

Knowledge of the limits atmospheric turbulence imposes on a free-space communication channel is very important for designing an optical system operating in such an environment. In this letter, we have experimentally characterized the effects of thin-phase turbulence over a range of $\ell = -5$ to $\ell= +5$, and verified that turbulence degrades the mode quality independent of input mode number. This result indicates that a system implementing adaptive optics to reduce the effects of turbulence can operate independently of the communications channel. The experimental data presented also indicates the potential working range of a free-space OAM channel and the expected cross-talk for such a system. We expect that our study provides useful information for the construction of practical quantum key distribution systems using OAM modes \cite{Boyd2011a}.

We acknowledge Dan Gauthier and Jonathan Leach for helpful discussions. Our work was primarily supported by the DARPA InPho program through the US Army Research Office award W911NF-10-1-0395. MPJL was further supported by European collaboration EC FP7 255914, PHORBITECH, and MJP is supported by the Royal Society. 


\bibliographystyle{ol.bst}

\end{document}